\documentclass[useAMS,usenatbib]{mn2e}
\usepackage{graphicx}
\usepackage{rotating,times,pictex,graphicx,latexsym}
\usepackage{color}
\usepackage{longtable,amsmath}
\usepackage{lscape}
\usepackage{array}
\usepackage{lipsum}
\newcolumntype{C}[1]{>{\centering\arraybackslash}p{1.1cm}}

\newcommand{\hii}{H{\sc ii}}

\title[New Compact Star Cluster Candidates]{New Compact Star Cluster Candidates in the Galactic Plane} 

\author[Froebrich, D.]{D.~Froebrich$^{1}$\thanks{E-mail:
df@star.kent.ac.uk}\\ $^1$ Centre for Astrophysics and Planetary Science,
University of Kent, Canterbury, CT2 7NH, UK } 

\begin{document}

\date{Received sooner; accepted later}
\pagerange{\pageref{firstpage}--\pageref{lastpage}} \pubyear{2011}
\maketitle

\label{firstpage}

\begin{abstract}

The sample of known star clusters, the fundamental building blocks of galaxies,
in the Milky Way is still extremely incomplete for objects beyond a distance of
1\,--\,2\,kpc. Many of the more distant and young clusters are compact and
hidden behind large amounts of extinction. We thus utilised the deep high
resolution near infrared surveys UGPS and VVV to uncover so far unknown compact
clusters and to analyse their properties. 

Images of all objects in the area covered by these two surveys, and which are
listed as Galaxy in SIMBAD have been inspected and 125 so far unknown stellar
clusters and candidate clusters have been identified. Based on the frequent
associations with star formation indicators (nebulosities, IRAS sources, \hii\
regions, masers) we find that the typical cluster in our sample is young, at
distances between 1\,--\,10\,kpc and has a typical apparent radius of
25\arcsec. We suggest more systematic searches e.g. at all positions of 2MASS
extended sources to increase the completeness of the known cluster sample
beyond distances of 2\,kpc.

\end{abstract}


\section{Introduction}

The vast majority of stars form in clusters \citep{Lada2003}, and most of these
clusters dissolve on timescales of 10s of Myrs \citep{2006MNRAS.373..752G} to
form the field star population. Thus, star clusters are the building blocks of
galaxies and their study is intimately linked to understanding star formation and
in particular the formation of more massive stars which are almost exclusively
formed in clusters \citep{2012MNRAS.424.3037G}, with potential exceptions such
as discussed e.g. in \citet{2013ApJ...768...66O}.

Establishing large and complete cluster samples is vital if we for example want
to understand the details of the cluster destruction/dissolution processes. In
particular the study of the impact of tidal forces from the galactic potential,
requires well characterised and bias free samples of clusters at a range of
galactocentric radii. The current most complete lists of clusters and their
properties contain a few thousand objects (e.g. about 2200 in the up to date
online version\footnote{http://www.wilton.unifei.edu.br/ocdb/} of
\citet{2002A&A...389..871D} and about 3000 in \citet{2013A&A...558A..53K}, plus
a few hundred more in the additions by \citet{2014A&A...568A..51S} and
\citet{2015A&A...581A..39S}. The latter paper also finds that our sample of
known clusters is only complete to a distance of 1.8\,kpc from the Sun, with
still many old clusters missing at even shorter distances
\citep{2014A&A...568A..51S} and basically no clusters are known at distances
further away than 5\,kpc - see Fig.\,7 in \citet{2014A&A...568A..51S}.
\citet{2013A&A...558A..53K} also show that the projected surface number density
of clusters at the position of the Sun is about 130 clusters per square
kiloparsec in the disk. This implies that there should be of the order of
2\,--\,3\,$\times 10^4$ clusters in the Milky Way. Hence, only about 10\,\% of
all clusters are currently catalogued and an even smaller fraction has its
parameters determined accurately.

Many of the more distant undiscovered clusters are young and thus potentially
associated with large amounts of extinction and projected against the Galactic
plane. Furthermore, with typical sizes of about one parsec, these clusters will
be of an apparent angular size of less than one arcminute for distances of
about 3\,kpc or higher. Hence, to discover and potentially analyse clusters
missing in our current sample, deep high resolution infrared surveys are going
to be of importance. Extensive searches in 2MASS \citep{2006AJ....131.1163S}
have already uncovered a huge new population of new clusters and candidates
(e.g. \citet{2003A&A...400..533D}, \citet{2003A&A...404..223B},
\citet{2007MNRAS.374..399F}, \citet{2010AstL...36...75G}). But the new deeper
and higher resolution surveys such as UGPS \citep{Lucas2008} and VISTA VVV
\citep{Minniti2010} will be able to uncover so far unrecognised objects which
are more compact and fainter. Some searches for new clusters in these surveys
have already been done (e.g. \citet{2012A&A...542A...3S},
\citet{2011A&A...532A.131B}, Lucas et al. (2017) in prep.). However, many
compact objects might have been overlooked. We thus aim to search these surveys
at the positions of known extended, but potentially miss-classified objects and
characterise them.

Our paper is structured as follows: The target selection and data analysis
procedures are explained in Sect.\,\ref{dataandanalysis}. We present the newly
discovered clusters and cluster candidates and their properties in
Sect.\,\ref{results}.

\section{Data analysis}\label{dataandanalysis}

\subsection{Target Selection}

The fact that a large fraction clusters at distances above about 2\,kpc has not
yet been discovered, is to some extent, but not entirely due to extinction in
the Galactic plane and the large number of foreground and background stars in
low latitude fields. One additional problem is spatial resolution. Assuming a
typical cluster size of 0.5\,pc, then at 5\,kpc the cluster would appear only
to have an extend of 20\arcsec. Thus, even if there are many bright, detectable
members, typical surveys such as 2MASS are not able to resolve all of the stars
into individual point sources and the cluster will only be detected as an
extended object instead. This is in particular true for young, embedded
clusters which are typically also surrounded by reflection/emission nebulae. As
young clusters constitute the majority of all clusters in the Galaxy (e.g.
\citet{Lada2003}, \citet{2014MNRAS.444..290B}), a large fraction of them might
be detectable even in 2MASS but only as extended sources, which in many cases
might have been classified (wrongly) as potential galaxies.

Hence, in order to identify potentially miss-classified star clusters, we used
the SIMBAD\footnote{\tt http://simbad.u-strasbg.fr/simbad/} database. We
extracted all objects from SIMBAD that were listed as Galaxy ({\tt otype =
'G'}). We planed to analyse the NIR data from the deep high resolution UGPS and
VISTA VVV surveys for these objects to establish their true nature. Thus, we
limited the selection to objects with a $K$-band detection (i.e. {\tt $K$mag <
20.0}) in SIMBAD. This ensures we do not select extended objects only detected
at mid or far-infrared wavelengths, but most likely sources with even an
extended 2MASS source counterpart. Note that the limiting magnitude for extended
sources in 2MASS is about $K$\,=\,14\,mag \citep{2000AJ....119.2498J}.

We also limited the search in SIMBAD to regions overlapping the UGPS and VVV
survey fields. The borders of the surveys where chosen as listed below. Please
note that the UGPS has its images orientated along the RA/DEC coordinates, while
the VVV data has an image orientation along the Galactic coordinate system.
Thus, the applied restrictions for UGPS (especially in latitude) are not the
exact survey limits.

\begin{itemize}

\item For UGPS: $|b| < 2^\circ$ and $359^\circ < l < 15^\circ$
\item For UGPS: $|b| < 5^\circ$ and $15^\circ < l < 107^\circ$
\item For UGPS: $|b| < 5^\circ$ and $141^\circ < l < 230^\circ$
\item For VVV: $|b| < 2^\circ$ and $295^\circ < l < 350^\circ$ (disk)
\item For VVV: $-10^\circ < b < 5^\circ$ and $350^\circ < l < 10^\circ$ (bulge)
\end{itemize}

Thus, our selection covers a total area that we searched for potential
miss-identified star clusters of about 1874 square degrees in UGPS and 520 square
degrees in VVV. There is some overlap of 44 square degrees between UGPS and VVV
near the Galactic Centre, where both surveys have available data. Thus, the
total area surveyed is 2350 square degrees. 

In total we find 4387 objects which pass our selection criteria, i.e. are
listed as Galaxy in SIMBAD, have a detected $K$-band counterpart of at least
20$^{\rm th}$ magnitude and are within the UGPS or VVV footprint.

\subsection{NIR data}

To investigate the high resolution UGPS and VVV images of the above selected 4387
objects in detail we have used the WSA\footnote{\tt
http://wsa.roe.ac.uk//index.html} and VSA\footnote{\tt
http://horus.roe.ac.uk/vsa/index.html} databases to extract $JHK(s)$ images for
each target. We extracted image cutouts with a size of 3\arcmin$\times$3\arcmin\
around each object. For some targets no images or not all three images were
available. This can be caused by the object either being close to the survey
boundary (which is not straight) or some images are not (yet) included in the
latest available data release due to not passing quality thresholds. Note that
we used DR10 of the UGPS and DR4 for the VVV data.

In total 3409 (77.7\,\%) of our targets have all three NIR images available in
either UGPS or VVV. We have combined these images into $JHK(s)$ colour
composites and visually inspected all of them to select all objects that could
be real star clusters. For each target the colour composite images have been
looked at in various zoom levels as well as different contrast settings to
ensure all potential star clusters can be identified. 

\begin{figure*}
\centering
\includegraphics[angle=0,width=\textwidth]{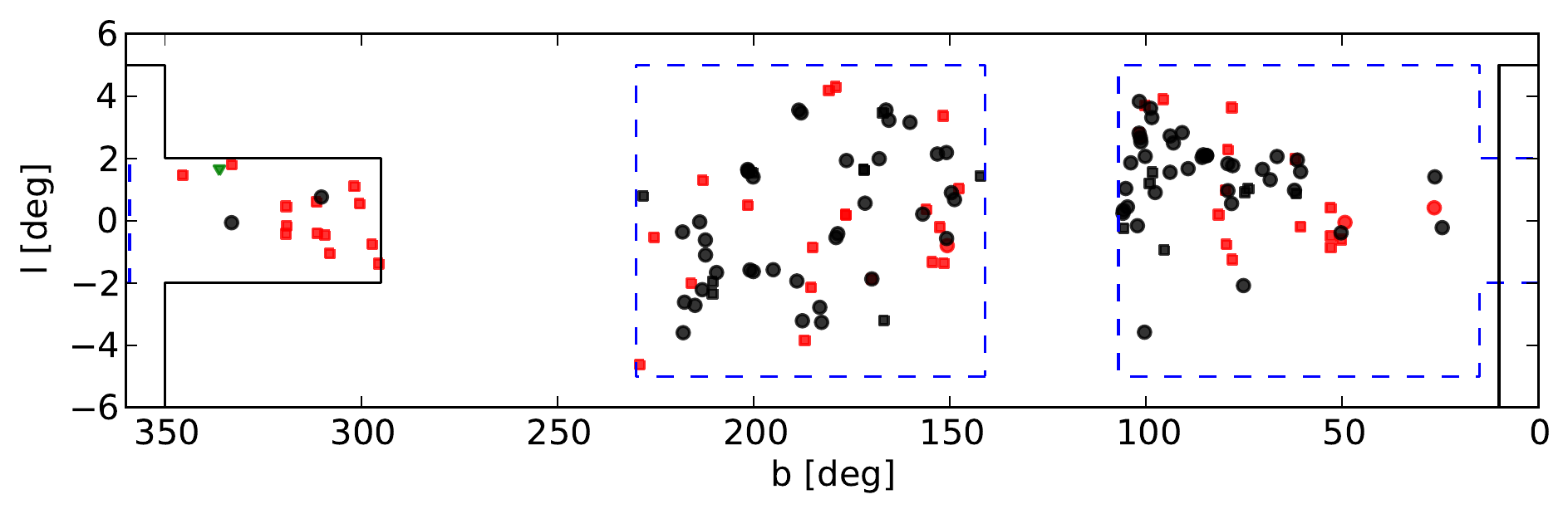} 

\caption{Distribution along the Galactic plane of all new clusters and candidate
objects. The symbols indicate the following: black circles - new clusters;
black square - confirmed clusters; red circle - new cluster candidates; red
squares - confirmed cluster candidates; green triangle - unknown object; The
area outlined by the black solid line indicated the VVV survey coverage, the
area outlined by the blue dashed line indicates the UGPS coverage.
\label{l_b_dist}}

\end{figure*}

\subsection{Star Cluster Identification}

From all the $JHK$ images we selected every object that visually could either be
identified as a group or cluster of stars or a nebulous young stellar object. In
other words we removed all the obvious galaxies from the target list. An
investigation of the properties of all these galaxies as well as potential new
galaxy clusters is planned, but outside the scope of this paper. Other objects
we removed were clearly associated with extended emission from edges of \hii\
regions and not stellar clusters. We also removed double entries, i.e. objects
that had two entries as Galaxy in SIMBAD very close to each other, in the same
cluster. Finally, all remaining objects where checked if they have an entry as a
known star cluster in SIMBAD.

The above process resulted in the following:

\begin{itemize}

\item There are 125 candidates for so far unknown/unpublished groups or
clusters of (mostly) young stars; 

\item There are 19 objects that are known cluster candidates; There are 84 known
star clusters which are not analysed any further in this paper; 

\item There is one object that cannot be reliably identified as either a star
cluster candidate or background galaxy (see discussion below).

\end{itemize}

Images of all the new star cluster candidate objects have been inspected a
second time in detail to determine some of the cluster properties. We manually
estimated the central position of the cluster, as well as its apparent radius
($R_{Cl}$) and also approximately counted manually the apparent number of stars
visible in the NIR images that seem to belong to the cluster. 

Based on this inspection, we further divided the sample of 125 objects into
clear cluster candidates (star clusters, hereafter), with an obvious overdensity
of NIR detectable stars and less clear cluster candidate objects (candidate
clusters hereafter). The latter mostly encompasses very nebulous objects and
small/less populous groups of young stars. This split has resulted in 77 new
clusters and 48 cluster candidates. Similarly to the new clusters, we also
classified the 19 known cluster candidates. In total 16 of these we judge to be
clusters, the remaining 3 fall into the candidate cluster category.

Finally, all the 125 candidate objects, as well as the 19 known cluster
candidates have been checked if they are associated with an \hii\ region or a
maser, and if the NIR images show a detectable nebulosity. We also cross-matched
all objects with IRAS and MSX sources and in particular objects in the Red MSX
Source (RMS) survey \citep{2013ApJS..208...11L} database\footnote{\tt
http://rms.leeds.ac.uk/cgi-bin/public/RMS$\_$DATABASE.cgi}. The latter objects
have in most cases a known distance determined from radial velocities
\citep{2014MNRAS.437.1791U}, which we have extracted. We list all the properties
of the objects in Table\,\ref{new_clusters_data} in the Appendix. Note that of
the 125 candidate objects 26 have been previously discovered in UGPS data by an
alternative search method, and will be published in Lucas et al. (2017, in
prep.). These are 24 of our 77 new clusters and 2 of our 48 new cluster
candidates. We refrain from listing the properties of these objects in
Table\,\ref{new_clusters_data} and also do not show their images in
Tables\,\ref{new_clusters_images} and \ref{new_cluster_candidates_images} as
they will be presented in Lucas et al. (2017, in prep.). They are, however,
included in our general analysis of their properties in Sect.\,\ref{results} and
are shown in Fig.\,\ref{l_b_dist}.

\subsection{Photometry and Cluster Membership}\label{phot_membership}

To investigate the cluster candidates and their members in more detail, we have
downloaded the full point source catalogues within 5' of each object from the
WSA and VSA databases. We extract the $JHK(s)$ magnitudes (using {\tt
AperMag3}) as well as the {\tt pstar} parameter for each star and only include
stars with {\tt pstar} $> 0.9$ and detections in all three filters in our
subsequent analysis. See \citet{Lucas2008} and \citet{Minniti2010} for a
detailed descriptions of how the magnitudes are measured and the {\tt pstar}
value for each star is determined.

For each cluster candidate we calculate the number of additional stars
($N_{Cl}$) within $R_{Cl}$ compared to the surrounding area (outside $R_{Cl}$,
within 5'). This number gives a measure of how significant an overdensity of
stars the cluster represents compared to the surrounding field i.e. estimates
the number of NIR detectable cluster members with reliable photometry. In
regions of extended large dark clouds and increased extinction these estimates
can of course be erroneous. Furthermore, even in these high resolution surveys,
many stars are not resolved and thus will not have a sufficiently accurate
brightness measurement to be included in the analysis. Especially in young and
very embedded clusters, which make up a large fraction of our objects (see
below), many stars are not detected at the shorter ($J$-band) wavelengths. Thus,
some clusters even show a negative number of members compared to the background
population despite a clear cluster being visible in the $K$-band images (e.g.
Cl\,019, Cl\,043 and in particular Cl\,062). 

We also investigated NIR colour-magnitude ($J$-$K$ vs. $K$) diagrams (CMDs) and
colour-colour ($H$-$K$ vs $J$-$H$) diagrams (CCDs) of stars in the cluster area
compared to the control field around them. However, due to the limited
photometry available, and the generally small number of stars, it has been
proven impossible to use well established photometric decontamination techniques
(e.g. following \citet{2007MNRAS.377.1301B}, \citet{2010MNRAS.409.1281F} or
\citet{2014MNRAS.444..290B}) to reliably identify the most likely cluster
members in order to fit isochrones to these diagrams. As there is basically
nothing to be gained from these diagrams we refrain from showing them. It might
be possible with careful psf-photometry of the UGPS/VVV images to improve the
photometry, but this is beyond the scope of this paper.

\begin{figure*}
\centering
\includegraphics[angle=0,width=\columnwidth]{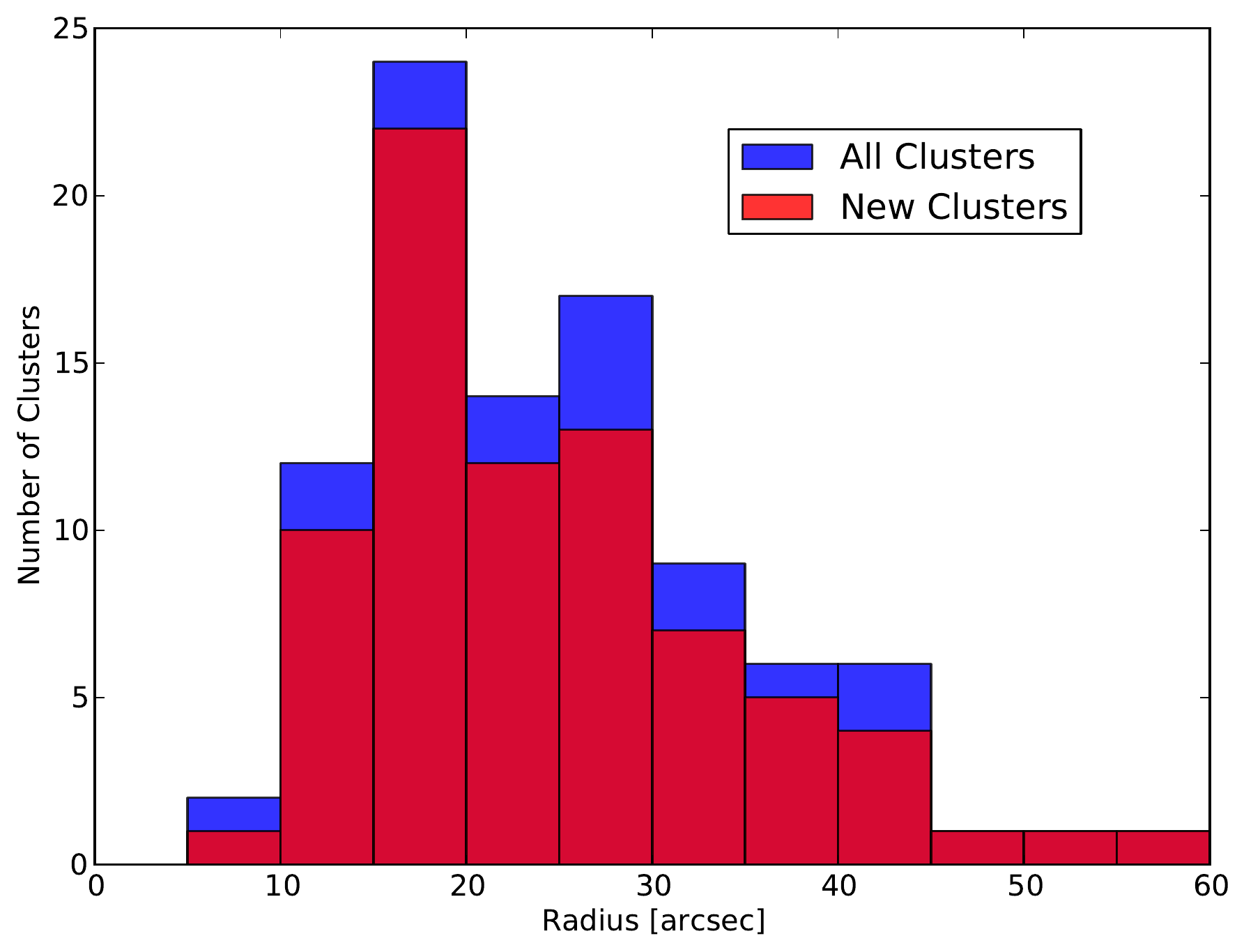} \hfill
\includegraphics[angle=0,width=\columnwidth]{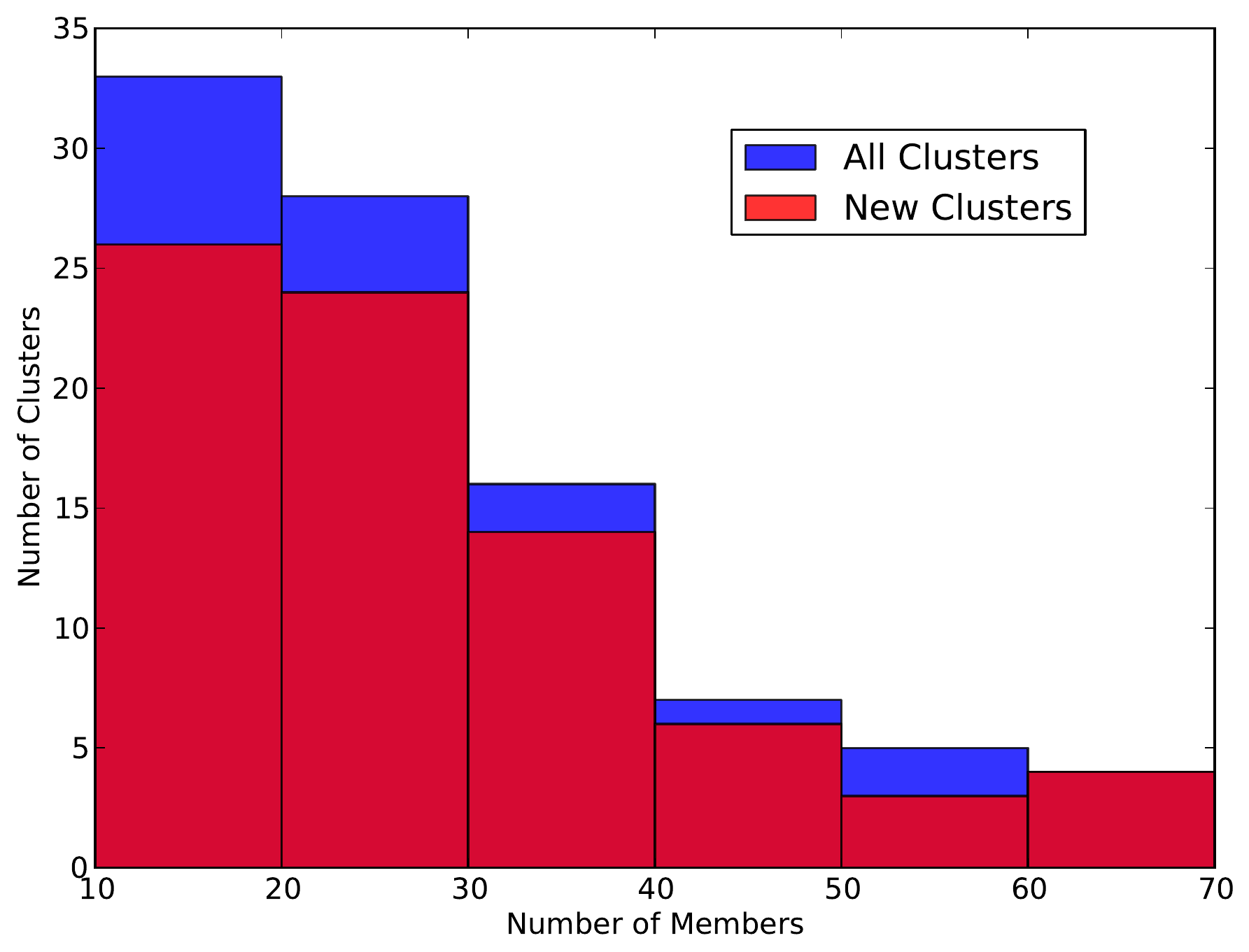} 

\caption{Distribution of the apparent radii (left panel) and number of manually
detected cluster members (right panel) of the clusters discovered in our sample.
The red histogram represents the newly discovered clusters, while the blue
histogram also includes the new clusters identified amongst the known cluster
candidates. Both histograms include all objects, including the ones that are
included in the list from Lucas et al. (2017, in prep.). \label{distributions}}

\end{figure*}

\section{Results and Discussion}\label{results}

\subsection{Distribution of New Clusters and Candidates}

In total we have identified 77 new clusters, 53 of which are not included in
the list of Lucas et al. (2017, in prep.). The images of these clusters are
shown in Table\,\ref{new_clusters_images} in Appendix\,\ref{app_images} and all
the determined  properties are listed in Table\,\ref{new_clusters_data}.  We
also identify 48 objects as candidate clusters or groups of stars, 46 of which
are not included in the list of Lucas et al. (2017, in prep.). The images of
these candidates are also shown in Table\,\ref{new_cluster_candidates_images}
in Appendix\,\ref{app_images} and all the properties are listed in
Table\,\ref{new_clusters_data}. 

We show the general distribution along the Galactic plane of all the newly
discovered objects in Fig.\,\ref{l_b_dist}. The figure shows that most of the
newly discovered clusters are away from the Galactic Centre. Only two of the new
clusters are in the area covered by VVV, the remainder are in the UGPS area.
This can to some extend be attributed to the density of SIMBAD objects
classified as Galaxy in the different fields. While the disk part of VVV has 0.6
'G' type objects per square degree, the equivalent UGPS field has 1.0 such
objects per square degree and the UGPS field in the outer Galaxy has 3.4 objects
per square degree. The latter is certainly caused by the lower extinction and
thus many more real galaxies in the sample. The differences between the two UGPS
and VVV inner disk fields are very minor when considering the actual extend. The
number of new clusters and candidates found in the equivalent area of the inner
disk in UGPS is almost identical to the number found in the VVV area. Hence,
most of the new clusters are found more than 60 degrees away from the Galactic
Centre.

Amongst the known cluster candidates in our list (most are from
\citet{2012A&A...542A...3S}, one from \citet{2005A&A...435..107I} and one from
\citet{2015MNRAS.450.4150C}), we identify 16 as real clusters. In case of the
remaining 3 objects no clear verdict can be reached from our data. Thus, these
objects remain cluster candidates. All their images are shown in
Table\,\ref{known_cluster_candidates_images} in Appendix\,\ref{app_images} and
all the properties are listed in Table\,\ref{new_clusters_data}.  In particular
object Cl\,142, known as Camargo\,442 \citep{2015MNRAS.450.4150C}, seems not to
be a real cluster. Instead the $JHK$ images suggest that this is just extended
nebulosity surrounding a single star and not a cluster.

One of our objects (Cl\,145) is of an unclear nature (see
Fig.\,\ref{unknown_object_images} in Appendix\,\ref{app_images}). It is situated
in a very crowded field near the Galactic Centre. A detailed inspection of the
$JHK$ image seems to suggest an increased number of fainter stars towards the
centre as well as nebulosity. On the other hand, it is not clearly identifiable
as a background galaxy. The 2MASS images of the globular cluster FSR\,1735
\citep{2007MNRAS.374..399F} appear somehow similar to this object, hence there
is still a possibility that this is a compact, distance rich cluster and not a
background galaxy.

\subsection{Properties of New Clusters and Candidates}

The properties of all new clusters, new cluster candidates and known candidates
in our sample are summarised in Table\,\ref{new_clusters_data}. The table is
organised by object group (new clusters, new candidates, etc.) and lists the
object identifier, the coordinates (J2000 and galactic), the cross
identification IDs from IRAS and MSX, associations with known \hii\ regions,
masers and nebulosities (identified in the $JHK$ images), the distances
obtained from associated RMS sources, the apparent radii as well as the number
of potential members determined manually from the images and automatically from
the photometric catalogues.

Of all our objects, 88.3\,\% have an association with an IRAS source. This
percentage is basically identical for the new clusters and the new cluster
candidates. Similarly, 25.5\,\% of all objects are associated with an MSX source
with identical percentages for the different groups.

The fraction of objects that are associated with star formation indicators such
as \hii\ regions and masers are also investigated. In total 27.6\,\% of objects
are associated with \hii\ regions and 10.3\,\% with masers. There are slight
differences in the percentages between the new clusters and the new candidates.
While about 1/3 of the new clusters are associated with \hii\ regions, only 1/5
of the new cluster candidates have such an association. This could be due to
the fact the \hii\ regions are indicators of slightly more massive and evolved
stars, thus increasing the likelihood of detecting more stars in the $JHK$
images and thus their classification as a new cluster. We find that 1/10 of the
new clusters and 1/6 of the new cluster candidates have associations with known
masers. As masers are indicators of more embedded younger massive protostars,
it is understandable that due to the increased extinction and younger age of
these objects it is more difficult to identify a sufficient number of members
in the $JHK$ images to classify the objects as new clusters.

In total 77.9\,\% of all objects are associated with detectable nebulous NIR
emission. Again, there is some variation of this fraction between the new
clusters (74.0\,\%) and the new cluster candidates (87.5\,\%). Similarly to the
association with \hii\ regions and masers, this difference can be explained by
extinction. Increased amounts of dust tend to lead to more detectable nebulosity
and at the same time decreased detectability of cluster members and thus, more
likely a classification of an object as cluster candidate.

We show the distribution of apparent radii measured for all clusters (known
candidates and new clusters) in the left panel of Fig.\,\ref{distributions}.
Most clusters have apparent radii between 10\arcsec\ and 45\arcsec\ with a wide
distribution. The median radius of the distribution is 23\arcsec. Thus, all the
clusters are quite compact, caused by the selection of the objects as
miss-classified galaxies. Objects much more extended than this would most likely
have been classified correctly as clusters even with lower resolution 2MASS
data. For comparison, the MWSC list by \citet{2013A&A...558A..53K}, including
the additions by \citet{2014A&A...568A..51S} and \citet{2015A&A...581A..39S},
does only contain very few objects with a core radius of less than 36\arcsec,
and no objects at all with an apparent radius smaller than 60\arcsec. Given the
large number of compact clusters identified in this work, compared to the small
number of known objects in the current up to date literature shows the potential
for future discoveries in these high resolution infrared surveys.

We display the distribution of the manually estimated cluster members for all
clusters (known candidates and new clusters) in the right panel of
Fig.\,\ref{distributions}. As to be expected there are significantly more
clusters with a smaller number of members. The median number of manually
detected members is 23 stars. The distribution of automatically detected members
in the clusters looks similar, just with a smaller number of clusters in each
bin. This is caused, as explained in Sect.\,\ref{phot_membership}, by the fact
that in regions with increased extinction the automatic determination of the
number of members fails, as background stars to the cluster are not detected in
the cluster field.

As can be seen in Table\,\ref{new_clusters_data}, the range of distances for the
clusters with associations to RMS objects is very large. The distances range
from just above 1\,kpc to slightly more than 10\,kpc, with a median value of
about 6\,kpc. These clusters are hence typically much further away than the
objects listed in the MWSC list, which have typical distances of
1.5\,--\,2.5\,kpc. The typical distances combined with the median apparent
radius of the clusters, lead to a typical radius of 0.7\,pc for the clusters in
our sample. However, only a fraction of 18\,\% of the new clusters have a
distance estimate. One could try to fit isochrones to potential cluster members
in order to obtain further distance estimates for the remainder of the sample.
However, as discussed in Sect.\,\ref{phot_membership}, it has proven not
possible to do with the currently available pipeline aperture photometry from
UGPS/VVV. This is mostly caused by the compactness of the new clusters and the
in part high extinction, which prevents detection and/or reliable photometry in
the $J$-band.

\section{Conclusions}

In order to identify potentially undetected star cluster (candidates) in the
Galactic plane, we manually inspected images from the UGPS and VVV surveys for
all objects classified as 'Galaxy' in SIMBAD to identify miss-classified star
clusters. A total of 4387 objects where initially selected and $JHK$ images for
3409 of those where available and investigated in detail in an area of 2350
square degrees over both surveys. 

Amongst the investigated objects we identified 125 so far unknown cluster
candidates, 19 known cluster candidates and 1 object of unclear nature. We
performed a detailed inspection of the so far unknown objects and identified 77
new clusters and 48 cluster candidates. All but three of the already known
cluster candidates can be confirmed as clusters with the utilised higher
resolution UGPS and VVV data.

A large fraction of the newly identified clusters and candidates are associated
with star formation indicators. Overall, about 80\,\% have detectable NIR
nebulosities, about 90\,\% are associated with IRAS sources, 25\,\% with MSX
sources and \hii\ regions and 10\,\% with masers. We extracted the distances of
the associated MSX sources and find that they range from 1 to 10\,kpc, with a
median of about 6\,kpc. The typical apparent radius of the new clusters is
about 25\arcsec\, corresponding to about 0.7\,pc at the typical distance. Thus,
our newly identified clusters are young, apparently compact clusters at
distances of several kiloparsec. Detailed cluster member identification and
isochrone fitting for the new objects cannot be performed as the currently
available photometry is not of sufficient quality due to the compactness of the
detected clusters. PSF photometry of the original survey data might in future
allow such investigations.

The large number of new clusters detected by our search is promising for future
work. While the discovered objects are in no way numerous enough to account for
the incompleteness in our cluster samples beyond about 1\,--\,2\,kpc, they show
that many of these missing clusters can be discovered in the available datasets.
A more systematic search in UGPS/VVV, e.g. at all positions of extended 2MASS
sources or near groups of UGPS/VVV objects with low {\tt pstar} values, should
reveal a substantial number of so far uncatalogued clusters.

\section*{acknowledgements}

The author would like to thank J.\,Urquhart for providing the RMS based
distances reported in the paper. He also acknowledges the comments on some of
the sources provided by D.\,Minniti, P.W.\,Lucas and T.\,Gledhill.

\bibliographystyle{mn2e}
\bibliography{biblio}

\clearpage
\newpage

\onecolumn

\begin{appendix}

\setlength\LTcapwidth{\textwidth}

\clearpage
\newpage

\begin{landscape}

\section{Cluster Data Table}

\begin{center} \begin{longtable}{|c|c|c|c|c|c|c|c|c|c|c|c|c|c|} \caption{This
table lists the properties of all clusters and candidates that are not included
in the list of Lucas et al. (2017, in prep). The table is organised by object
group (new clusters, new candidates, etc.) and lists the object identifier, the
coordinates (J2000 and galactic), the cross identification IDs from IRAS and
MSX, associations with known \hii\ regions, masers and nebulosities (identified
in the $JHK$ images), the distances obtained from associated RMS sources, the
apparent radii as well as the number of potential members determined manually
from the images and automatically from the photometric catalogues.
\label{new_clusters_data}} \\

\hline \multicolumn{1}{|c|}{\textbf{Name}} &
\multicolumn{4}{c|}{\textbf{Coordinates}} &
\multicolumn{2}{c|}{\textbf{Cross-ID}} &
\multicolumn{3}{c|}{\textbf{Association}} &
\multicolumn{1}{c|}{\textbf{Distance}}  &
\multicolumn{1}{c|}{\textbf{Radius}}  &
\multicolumn{2}{c|}{\textbf{Number}} \\ 
\multicolumn{1}{|c|}{\textbf{}} &
\multicolumn{2}{c|}{\textbf{RA/DEC (J2000)}} &
\multicolumn{2}{c|}{\textbf{Galactic (l,b)}} &
\multicolumn{2}{c|}{\textbf{}} &
\multicolumn{3}{c|}{\textbf{with}} &
\multicolumn{1}{c|}{\textbf{}}  &
\multicolumn{1}{c|}{\textbf{}}  &
\multicolumn{2}{c|}{\textbf{of Stars}} \\ 
\multicolumn{1}{|c|}{\textbf{}} &
\multicolumn{1}{c|}{\textbf{[h:m:s]}} &
\multicolumn{1}{c|}{\textbf{[$^\circ$:$'$:$''$]}} &
\multicolumn{2}{c|}{\textbf{[deg]}} &
\multicolumn{1}{c|}{\textbf{IRAS}} &
\multicolumn{1}{c|}{\textbf{MSX}} &
\multicolumn{1}{c|}{\textbf{\hii}} &
\multicolumn{1}{c|}{\textbf{Maser}} &
\multicolumn{1}{c|}{\textbf{Nebula}} &
\multicolumn{1}{c|}{\textbf{[kpc]}}  &
\multicolumn{1}{c|}{\textbf{[arcsec]}}  &
\multicolumn{1}{c|}{\textbf{Man.}} &
\multicolumn{1}{c|}{\textbf{Aut.}} \\ \hline  
\endfirsthead

\multicolumn{14}{c}%
{{\bfseries \tablename\ \thetable{} -- continued from previous page}} \\

\hline \multicolumn{1}{|c|}{\textbf{Name}} &
\multicolumn{4}{c|}{\textbf{Coordinates}} &
\multicolumn{2}{c|}{\textbf{Cross-ID}} &
\multicolumn{3}{c|}{\textbf{Association}} &
\multicolumn{1}{c|}{\textbf{Distance}}  &
\multicolumn{1}{c|}{\textbf{Radius}}  &
\multicolumn{2}{c|}{\textbf{Number}} \\ 
\multicolumn{1}{|c|}{\textbf{}} &
\multicolumn{2}{c|}{\textbf{RA/DEC (J2000)}} &
\multicolumn{2}{c|}{\textbf{Galactic (l,b)}} &
\multicolumn{2}{c|}{\textbf{}} &
\multicolumn{3}{c|}{\textbf{with}} &
\multicolumn{1}{c|}{\textbf{}}  &
\multicolumn{1}{c|}{\textbf{}}  &
\multicolumn{2}{c|}{\textbf{of Stars}} \\ 
\multicolumn{1}{|c|}{\textbf{}} &
\multicolumn{1}{c|}{\textbf{[h:m:s]}} &
\multicolumn{1}{c|}{\textbf{[$^\circ$:$'$:$''$]}} &
\multicolumn{2}{c|}{\textbf{[deg]}} &
\multicolumn{1}{c|}{\textbf{IRAS}} &
\multicolumn{1}{c|}{\textbf{MSX}} &
\multicolumn{1}{c|}{\textbf{\hii}} &
\multicolumn{1}{c|}{\textbf{Maser}} &
\multicolumn{1}{c|}{\textbf{Nebula}} &
\multicolumn{1}{c|}{\textbf{[kpc]}}  &
\multicolumn{1}{c|}{\textbf{[arcsec]}}  &
\multicolumn{1}{c|}{\textbf{Man.}} &
\multicolumn{1}{c|}{\textbf{Aut.}} \\ \hline  
\endhead

\multicolumn{14}{|r|}{{Continued on next page}} \\ \hline
\endfoot

\endlastfoot

\multicolumn{14}{|c|}{{\bf New Clusters}} \\ \hline
Cl\,001 & 04:01:26.4 & $+$53:43:17 & 148.799412 & $+$0.677165 & IRAS\,03575$+$5334 &          --         &  N  & N & N &   -- & 18 & 30 & 11 \\ \hline 
Cl\,002 & 04:05:56.6 & $+$51:27:05 & 150.814877 & $-$0.571217 &         --         &          --         &  N  & N & Y &   -- & 15 & 10 & -- \\ \hline 
Cl\,003 & 04:06:25.5 & $+$53:21:49 & 149.591151 & $+$0.900032 & IRAS\,04025$+$5313 &          --         &  N  & N & Y &   -- & 26 & 40 &  4 \\ \hline 
Cl\,004 & 04:18:32.6 & $+$53:26:03 & 150.859922 & $+$2.188007 & IRAS\,04146$+$5318 & G150.8602$+$02.1879 & Y/N & Y & N &   -- & 32 & 50 & 24 \\ \hline 
Cl\,005 & 04:29:00.1 & $+$51:45:23 & 153.171059 & $+$2.140904 & IRAS\,04251$+$5138 &          --         &  Y  & N & Y &   -- & 28 & 15 & 21 \\ \hline 
Cl\,006 & 04:35:58.9 & $+$47:43:03 & 156.895187 & $+$0.212232 & IRAS\,04322$+$4736 &          --         &  N  & N & N &   -- & 20 & 30 &  5 \\ \hline 
Cl\,007 & 05:01:39.7 & $+$47:07:22 & 160.144438 & $+$3.155929 & IRAS\,04579$+$4703 & G160.1452$+$03.1559 &  Y  & N & Y & 1.94 & 27 & 40 & -- \\ \hline 
Cl\,013 & 05:39:28.4 & $+$24:56:30 & 182.678494 & $-$3.265738 & IRAS\,05363$+$2454 &          --         &  Y  & Y & N &   -- & 10 & 15 &  4 \\ \hline 
Cl\,014 & 05:40:29.6 & $+$29:56:48 & 178.553723 & $-$0.418188 & IRAS\,05372$+$2955 &          --         &  N  & N & N &   -- & 16 & 10 &  3 \\ \hline 
Cl\,015 & 05:41:05.2 & $+$29:30:26 & 178.994673 & $-$0.541804 & IRAS\,05378$+$2928 &          --         &  N  & N & Y &   -- & 25 & 25 & 12 \\ \hline 
Cl\,016 & 05:42:20.6 & $+$24:48:36 & 183.134536 & $-$2.782362 & IRAS\,05393$+$2447 &          --         &  N  & N & N &   -- & 29 & 35 & 17 \\ \hline 
Cl\,017 & 05:44:23.9 & $+$33:03:51 & 176.341522 & $+$1.931052 & IRAS\,05411$+$3302 &          --         &  N  & N & Y &   -- & 56 & 30 & 25 \\ \hline 
Cl\,018 & 05:50:40.1 & $+$20:48:10 & 187.562049 & $-$3.214981 & IRAS\,05476$+$2047 &          --         &  N  & N & N &   -- & 26 & 15 & 14 \\ \hline 
Cl\,019 & 05:58:24.5 & $+$20:13:57 & 188.969785 & $-$1.937577 & IRAS\,05554$+$2013 & G188.9696$-$01.9380 &  N  & N & Y & 2.0  & 30 & 25 & -- \\ \hline 
Cl\,020 & 06:12:06.9 & $+$15:09:04 & 195.000702 & $-$1.575500 &       --         &          --         &  N  & N & Y &   -- & 19 & 15 & -- \\ \hline 
Cl\,021 & 06:16:31.8 & $+$23:47:23 & 187.902651 & $+$3.461314 & IRAS\,06134$+$2348 &          --         &  N  & N & Y &   -- & 15 & 30 &  6 \\ \hline 
Cl\,022 & 06:18:04.2 & $+$23:19:14 & 188.483899 & $+$3.551225 & IRAS\,06150$+$2320 &          --         &  N  & N & Y &   -- & 20 & 50 & 11 \\ \hline 
Cl\,023 & 06:21:47.7 & $+$10:39:23 & 200.078150 & $-$1.632706 & IRAS\,06190$+$1040 & G200.0789$-$01.6323 &  Y  & N & N & 5.78 & 40 & 70 & 20 \\ \hline 
Cl\,024 & 06:23:34.3 & $+$09:56:19 & 200.917363 & $-$1.583128 & IRAS\,06208$+$0957 & G200.9166$-$01.5827 & Y/N & N & Y & 5.54 & 34 & 40 & 10 \\ \hline 
Cl\,025 & 06:32:51.3 & $+$12:01:34 & 200.127573 & $+$1.404202 & IRAS\,06300$+$1203 &          --         &  N  & N & N &   -- & 19 & 30 &  6 \\ \hline 
Cl\,026 & 06:35:50.0 & $+$10:59:49 & 201.377469 & $+$1.578306 &          --         &          --         &  N  & N & N &   -- & 16 & 15 &  6 \\ \hline 
Cl\,027 & 06:35:56.1 & $+$11:00:19 & 201.381527 & $+$1.604296 & IRAS 06331$+$1102 &          --         &  N  & N & Y &   -- & 27 & 20 &  6 \\ \hline 
Cl\,028 & 06:36:19.3 & $+$10:54:26 & 201.512264 & $+$1.643583 & IRAS 06335$+$1057 &          --         &  N  & N & N &   -- & 17 & 30 & 15 \\ \hline 
Cl\,038 & 13:51:59.6 & $-$61:15:39 & 310.146375 & $+$0.759752 & IRAS\,13484$-$6100 &          --         &  N  & Y & Y &   -- & 30 & 20 & -- \\ \hline 
Cl\,039 & 16:18:57.4 & $-$50:23:59 & 333.029624 & $+$0.065123 &         --         &          --         &  N  & Y & Y &   -- & 26 & 20 & -- \\ \hline 
Cl\,041 & 18:36:47.1 & $-$07:35:41 &  24.507573 & $+$0.223496 & IRAS\,18340$-$0738 &          --         &  Y  & N & Y &   -- & 13 & 20 & -- \\ \hline 
Cl\,044 & 19:39:32.7 & $+$26:05:25 &  61.422407 & $+$1.947718 & IRAS\,19374$+$2558 &          --         &  N  & N & Y &   -- & 17 & 20 & -- \\ \hline 
Cl\,046 & 19:50:53.0 & $+$30:38:10 &  66.608918 & $+$2.060751 & IRAS\,19489$+$3030 &          --         &  N  & N & Y &   -- & 13 & 10 & -- \\ \hline 
Cl\,047 & 19:58:03.1 & $+$31:44:07 &  68.341812 & $+$1.313669 & IRAS\,19560$+$3135 & G068.3418$+$01.3138 &  Y  & N & Y &11.66 & 12 & 10 & -- \\ \hline 
Cl\,048 & 20:01:37.4 & $+$33:35:28 &  70.316479 & $+$1.648899 & IRAS\,19597$+$3327 & G070.3164$+$01.6493 & Y/N & N & Y & 7.42 & 17 & 20 & -- \\ \hline 
Cl\,049 & 20:21:55.1 & $+$39:59:46 &  77.900648 & $+$1.767362 &         --         & G077.8999$+$01.7678 &  Y  & N & Y & 1.4  & 23 & 10 & -- \\ \hline 
Cl\,050 & 20:25:25.3 & $+$41:03:19 &  79.150728 & $+$1.829256 & IRAS\,20236$+$4053 &          --         &  N  & N & Y &   -- & 15 & 20 & -- \\ \hline 
Cl\,051 & 20:27:58.8 & $+$39:32:09 &  78.196161 & $+$0.550244 & IRAS\,20261$+$3922 & G078.1952$+$00.5497 & Y/N & N & Y & 8.51 &  9 & 10 &  1 \\ \hline 
Cl\,053 & 20:29:46.8 & $+$35:31:42 &  75.154768 & $-$2.084232 & IRAS\,20278$+$3521 &          --         &  N  & Y & Y &   -- & 24 & 30 & -- \\ \hline 
Cl\,054 & 20:41:53.5 & $+$45:32:00 &  84.534232 & $+$2.089183 & IRAS\,20402$+$4521 &          --         &  N  & N & Y &   -- & 14 & 25 & -- \\ \hline 
Cl\,055 & 20:41:55.2 & $+$45:32:39 &  84.545836 & $+$2.091922 &         --         &          --         &  N  & N & Y &   -- & 12 & 15 & -- \\ \hline 
Cl\,056 & 20:44:53.1 & $+$46:14:15 &  85.410346 & $+$2.114474 & IRAS\,20431$+$4603 &          --         &  N  & N & Y &   -- & 13 & 15 & -- \\ \hline 
Cl\,057 & 20:46:17.2 & $+$46:24:35 &  85.695990 & $+$2.032407 & IRAS\,20446$+$4613 &          --         &  Y  & N & Y &   -- & 38 & 20 & -- \\ \hline 
Cl\,058 & 21:01:34.9 & $+$48:55:01 &  89.272880 & $+$1.670386 & IRAS\,20599$+$4843 & G089.2727$+$01.6703 &  Y  & N & Y & 8.06 & 23 & 30 & -- \\ \hline 
Cl\,059 & 21:02:21.8 & $+$50:48:35 &  90.777103 & $+$2.827686 & IRAS\,21007$+$5036 & G090.7764$+$02.8281 &  Y  & N & Y & 1.74 & 15 & 10 & -- \\ \hline 
Cl\,060 & 21:13:38.9 & $+$52:14:01 &  93.014939 & $+$2.495089 & IRAS\,21120$+$5201 & G093.0166$+$02.4953 &  Y  & N & N &   -- & 19 & 10 & -- \\ \hline 
Cl\,061 & 21:16:19.3 & $+$52:58:57 &  93.837221 & $+$2.720457 & IRAS\,21147$+$5246 &          --         &  N  & N & N &   -- & 16 & 20 &  2 \\ \hline 
Cl\,062 & 21:21:53.6 & $+$52:10:47 &  93.859904 & $+$1.554173 & IRAS\,21202$+$5157 & G093.8588$+$01.5551 &  Y  & N & Y & 6.36 & 33 & 70 & -- \\ \hline 
Cl\,063 & 21:36:46.0 & $+$56:38:56 &  98.506636 & $+$3.311011 & IRAS\,21351$+$5625 &          --         &  N  & N & Y &   -- & 17 & 20 & 13 \\ \hline 
Cl\,064 & 21:37:04.4 & $+$57:05:16 &  98.832037 & $+$3.609138 &         --         &          --         &  N  & N & N &   -- & 19 & 20 & -- \\ \hline 
Cl\,066 & 21:52:02.2 & $+$56:47:49 & 100.197892 & $+$2.064316 & IRAS\,21503$+$5633 & G100.1974$+$02.0643 & Y/N & N & Y & 5.85 & 29 & 40 & -- \\ \hline 
Cl\,067 & 21:52:30.6 & $+$59:06:52 & 101.703900 & $+$3.828645 & IRAS\,21509$+$5852 &          --         &  N  & N & N &   -- & 16 & 20 &  2 \\ \hline 
Cl\,068 & 21:56:15.1 & $+$57:50:42 & 101.296341 & $+$2.530582 & IRAS\,21545$+$5736 &          --         &  N  & N & Y &   -- & 23 & 15 & -- \\ \hline 
Cl\,069 & 21:56:27.3 & $+$58:01:50 & 101.432089 & $+$2.660077 & IRAS\,21548$+$5747 &          --         &  Y  & N & Y &   -- & 39 & 70 & 27 \\ \hline 
Cl\,070 & 21:57:44.6 & $+$58:21:06 & 101.763679 & $+$2.809178 & IRAS\,21561$+$5806 & G101.7639$+$02.8100 & Y/N & Y & Y & 7.81 & 32 & 25 & -- \\ \hline 
Cl\,071 & 22:13:16.6 & $+$56:12:14 & 102.192782 & $-$0.162377 & IRAS\,22114$+$5557 &          --         &  N  & N & Y &   -- & 13 & 15 & -- \\ \hline 
Cl\,072 & 22:15:08.7 & $+$58:49:16 & 103.875995 & $+$1.857624 & IRAS\,22134$+$5834 & G103.8744$+$01.8558 &  Y  & N & Y & 1.59 & 49 & 20 & 17 \\ \hline 
Cl\,075 & 22:26:57.4 & $+$58:49:41 & 105.163562 & $+$1.033268 & IRAS\,22251$+$5834 &          --         &  N  & N & Y &   -- & 14 & 10 & -- \\ \hline 
\multicolumn{14}{|c|}{{\bf New Cluster Candidates}} \\ \hline
Cl\,078 & 03:56:57.8 & $+$54:45:02 & 147.633127 & $+$1.033568 & IRAS\,03530$+$5436 &         --          &  N  & N & N &  --  & -- & -- & -- \\ \hline 
Cl\,079 & 04:05:47.0 & $+$50:25:04 & 151.486411 & $-$1.357451 & IRAS\,04020$+$5017 &         --          &  N  & N & N &  --  & -- & -- & -- \\ \hline 
Cl\,080 & 04:19:10.7 & $+$48:24:57 & 154.445183 & $-$1.325439 & IRAS\,04154$+$4817 &         --          &  N  & N & Y &  --  & -- & -- & -- \\ \hline 
Cl\,081 & 04:27:49.6 & $+$53:43:08 & 151.627347 & $+$3.367920 & IRAS\,04238$+$5336 &         --          &  N  & N & Y &  --  & -- & -- & -- \\ \hline 
Cl\,082 & 04:32:44.6 & $+$48:31:49 & 155.931246 & $+$0.365437 & IRAS\,04290$+$4825 &         --          &  N  & N & Y &  --  & -- & -- & -- \\ \hline 
Cl\,083 & 04:15:22.0 & $+$50:34:36 & 152.496404 & $-$0.205135 & IRAS\,04115$+$5027 &         --          &  N  & N & Y &  --  & -- & -- & -- \\ \hline 
Cl\,084 & 05:11:22.8 & $+$36:21:20 & 169.835550 & $-$1.878073 & IRAS\,05081$+$3616 &         --          &  N  & N & N &  --  & -- & -- & -- \\ \hline 
Cl\,085 & 05:37:42.0 & $+$32:00:48 & 176.486090 & $+$0.177157 &        --          &         --          &  N  & N & N &  --  & -- & -- & -- \\ \hline 
Cl\,086 & 05:37:50.8 & $+$32:00:40 & 176.504622 & $+$0.202220 &        --          &         --          &  N  & N & N &  --  & -- & -- & -- \\ \hline 
Cl\,087 & 05:47:04.3 & $+$20:59:41 & 186.965545 & $-$3.837873 &        --          &         --          &  N  & N & Y &  --  & -- & -- & -- \\ \hline 
Cl\,088 & 05:49:39.7 & $+$23:16:59 & 185.308898 & $-$2.146923 & IRAS\,05466$+$2316 &         --          &  N  & Y & Y &  --  & -- & -- & -- \\ \hline 
Cl\,089 & 05:53:43.6 & $+$24:14:45 & 184.954832 & $-$0.855921 & IRAS\,05506$+$2414 & G184.9551$-$00.8559 &  N  & N & Y & 3.66 & -- & -- & -- \\ \hline 
Cl\,090 & 06:00:33.4 & $+$31:56:44 & 179.037040 & $+$4.299981 & IRAS\,05573$+$3156 & G179.0380$+$04.3003 &  N  & N & Y & 1.1  & -- & -- & -- \\ \hline 
Cl\,091 & 06:04:10.1 & $+$30:20:24 & 180.820245 & $+$4.188061 & IRAS\,06009$+$3020 &         --          &  N  & N & Y &  --  & -- & -- & -- \\ \hline 
Cl\,092 & 06:31:58.7 & $+$10:27:48 & 201.414795 & $+$0.492048 & IRAS\,06292$+$1029 &         --          &  N  & N & Y &  --  & -- & -- & -- \\ \hline 
Cl\,093 & 06:49:40.2 & $-$03:32:52 & 215.889431 & $-$2.009736 & IRAS\,06471$-$0329 &         --          &  N  & N & Y &  --  & -- & -- & -- \\ \hline 
Cl\,095 & 07:04:21.6 & $-$16:23:20 & 228.993299 & $-$4.619786 & IRAS\,07020$-$1618 &         --          &  N  & N & Y &  --  & -- & -- & -- \\ \hline 
Cl\,096 & 07:12:24.5 & $-$11:15:34 & 225.326623 & $-$0.531525 & IRAS\,07100$-$1110 &         --          &  Y  & N & Y &  --  & -- & -- & -- \\ \hline 
Cl\,097 & 11:45:04.7 & $-$63:17:46 & 295.558094 & $-$1.378464 & IRAS\,11426$-$6301 &         --          &  N  & N & Y &  --  & -- & -- & -- \\ \hline 
Cl\,098 & 12:00:57.0 & $-$63:04:10 & 297.251695 & $-$0.755942 & IRAS\,11583$-$6247 &         --          &  N  & N & Y &  --  & -- & -- & -- \\ \hline 
Cl\,099 & 12:29:41.7 & $-$62:13:09 & 300.401305 & $+$0.545612 & IRAS\,12268$-$6156 &         --          &  N  & N & Y &  --  & -- & -- & -- \\ \hline 
Cl\,100 & 12:41:17.6 & $-$61:44:42 & 301.731465 & $+$1.103014 & IRAS\,12383$-$6128 &         --          &  N  & N & Y &  --  & -- & -- & -- \\ \hline 
Cl\,101 & 13:37:20.7 & $-$63:28:12 & 308.031655 & $-$1.053791 & IRAS\,13338$-$6312 &         --          &  N  & N & Y &  --  & -- & -- & -- \\ \hline 
Cl\,102 & 13:46:37.0 & $-$62:39:30 & 309.219610 & $+$0.462208 & IRAS\,13431$-$6224 &         --          &  N  & N & Y &  --  & -- & -- & -- \\ \hline 
Cl\,103 & 14:02:36.2 & $-$61:05:45 & 311.425369 & $+$0.596829 & IRAS\,13590$-$6051 &         --          &  Y  & N & Y &  --  & -- & -- & -- \\ \hline 
Cl\,104 & 14:02:52.8 & $-$62:07:23 & 311.177140 & $-$0.400493 & IRAS\,13592$-$6153 &         --          &  N  & N & Y &  --  & -- & -- & -- \\ \hline 
Cl\,105 & 14:59:29.2 & $-$58:20:10 & 319.089385 & $+$0.460397 & IRAS\,14556$-$5808 &         --          &  N  & N & Y &  --  & -- & -- & -- \\ \hline 
Cl\,106 & 15:00:35.0 & $-$58:58:10 & 318.915062 & $+$0.164954 &        --          &         --          &  Y  & Y & Y &  --  & -- & -- & -- \\ \hline 
Cl\,107 & 15:03:13.7 & $-$59:04:29 & 319.163190 & $-$0.420600 & IRAS\,14593$-$5852 &         --          &  Y  & N & Y &  --  & -- & -- & -- \\ \hline 
Cl\,108 & 16:10:38.7 & $-$49:05:59 & 332.955261 & $+$1.802102 & IRAS\,16069$-$4858 &         --          &  Y  & N & Y &  --  & -- & -- & -- \\ \hline 
Cl\,109 & 16:59:42.3 & $-$40:03:45 & 345.494751 & $+$1.466657 & IRAS\,16562$-$3959 &         --          &  N  & Y & Y &  --  & -- & -- & -- \\ \hline 
Cl\,110 & 19:25:57.9 & $+$15:03:02 &  50.222047 & $-$0.606428 & IRAS\,19236$+$1456 & G050.2213$-$00.6063 &  Y  & N & Y & 3.34 & -- & -- & -- \\ \hline 
Cl\,111 & 19:27:34.8 & $+$17:54:37 &  52.921318 & $+$0.414364 & IRAS\,19253$+$1748 & G052.9217$+$00.4142 &  N  & Y & Y & 5.06 & -- & -- & -- \\ \hline 
Cl\,112 & 19:30:54.7 & $+$17:28:44 &  52.922557 & $-$0.488824 & IRAS\,19286$+$1722 & G052.9221$-$00.4892 & Y/N & N & Y & 5.06 & -- & -- & -- \\ \hline 
Cl\,113 & 19:32:16.1 & $+$17:17:54 &  52.920171 & $-$0.859438 & IRAS\,19300$+$1711 & AGAL052.919$-$00.861&  N  & N & Y &  --  & -- & -- & -- \\ \hline 
Cl\,114 & 19:40:34.6 & $+$26:34:15 &  61.954815 & $+$1.982962 & IRAS\,19385$+$2627 &         --          &  N  & N & Y &  --  & -- & -- & -- \\ \hline 
Cl\,115 & 19:45:52.5 & $+$24:17:42 &  60.575275 & $-$0.186735 & IRAS\,19437$+$2410 & G060.5750$-$00.1861 &  Y  & N & Y & 7.48 & -- & -- & -- \\ \hline 
Cl\,116 & 20:14:25.9 & $+$41:13:37 &  78.123116 & $+$3.633933 & IRAS\,20126$+$4104 & G078.1224$+$03.6320 &  N  & Y & Y & 1.4  & -- & -- & -- \\ \hline 
Cl\,117 & 20:23:23.9 & $+$41:17:40 &  79.127561 & $+$2.278287 & IRAS\,20216$+$4107 & G079.1272$+$02.2782 &  Y  & N & Y & 1.4  & -- & -- & -- \\ \hline 
Cl\,118 & 20:30:50.9 & $+$41:02:30 &  79.736752 & $+$0.990351 & IRAS\,20290$+$4052 &         --          &  N  & Y & Y &  --  & -- & -- & -- \\ \hline 
Cl\,119 & 20:34:49.9 & $+$38:19:07 &  78.001769 & $-$1.243995 & IRAS\,20329$+$3808 &         --          &  N  & N & Y &  --  & -- & -- & -- \\ \hline 
Cl\,121 & 20:39:58.0 & $+$41:59:15 &  81.517594 & $+$0.192362 &        --          & G081.5168$+$00.1926 &  Y  & N & Y & 1.4  & -- & -- & -- \\ \hline 
Cl\,122 & 21:18:52.8 & $+$55:03:23 &  95.587783 & $+$3.898574 & IRAS \,1173$+$5450 &         --          &  N  & Y & Y &  --  & -- & -- & -- \\ \hline 
Cl\,123 & 21:44:04.8 & $+$58:01:46 & 100.162064 & $+$3.696814 & IRAS \,1425$+$5747 &         --          &  N  & N & Y &  --  & -- & -- & -- \\ \hline 
Cl\,124 & 21:55:41.8 & $+$57:58:31 & 101.318808 & $+$2.678575 & IRAS \,1540$+$5744 & G101.3193$+$02.6785 &  N  & N & Y & 6.24 & -- & -- & -- \\ \hline 
Cl\,125 & 21:57:05.5 & $+$58:17:54 & 101.663280 & $+$2.819535 & IRAS \,1554$+$5803 &         --          &  N  & N & N &  --  & -- & -- & -- \\ \hline 
\multicolumn{14}{|c|}{{\bf Confirmed Cluster Candidates}} \\ \hline
Cl\,126 & 03:27:31.4 & $+$58:19:22 & 142.245218 & $+$1.429872 & IRAS\,03235$+$5808 & G142.2446$+$01.4299 & Y/N & N & Y & 4.15 & 31 & 50 &  9 \\ \hline 
Cl\,127 & 04:56:55.2 & $+$37:57:17 & 166.813749 & $-$3.198304 & IRAS\,04535$+$3752 & G166.8141$-$03.1986 & Y/N & N & Y & 2.0  & 41 & 50 & 54 \\ \hline 
Cl\,128 & 05:25:40.6 & $+$41:41:53 & 167.059948 & $+$3.463957 & IRAS\,05221$+$4139 &         --          &  N  & N & Y &  --  & 42 & 40 & 13 \\ \hline 
Cl\,129 & 05:31:28.0 & $+$36:43:01 & 171.835302 & $+$1.644529 &        --          &         --          &  N  & N & N &  --  & 23 & 30 &  9 \\ \hline 
Cl\,130 & 05:31:28.7 & $+$36:41:54 & 171.852162 & $+$1.636293 &        --          &         --          &  N  & N & N &  --  & 19 & 20 &  5 \\ \hline 
Cl\,131 & 06:33:27.2 & $+$12:03:31 & 200.166170 & $+$1.549008 & IRAS\,06306$+$1205 &         --          &  N  & N & Y &  --  & 39 & 20 & 37 \\ \hline 
Cl\,132 & 06:38:36.6 & $+$01:07:20 & 210.469190 & $-$2.341123 & IRAS\,06360$+$0109 &         --          &  N  & N & N &  --  & 16 & 15 & -- \\ \hline 
Cl\,133 & 06:39:52.2 & $+$01:20:52 & 210.412543 & $-$1.957914 & IRAS\,06372$+$0123 &         --          &  N  & N & N &  --  & 25 & 15 &  3 \\ \hline 
Cl\,134 & 07:22:30.8 & $-$13:05:26 & 228.099152 & $+$0.796587 & IRAS\,07201$-$1259 &         --          &  N  & N & Y &  --  & 27 & 30 &  9 \\ \hline 
Cl\,135 & 19:44:23.5 & $+$25:48:42 &  61.720129 & $+$0.863272 & IRAS\,19423$+$2541 & G061.7201$+$00.8630 & Y/N & N & Y &13.99 & 14 & 20 & -- \\ \hline 
Cl\,136 & 20:13:34.2 & $+$36:15:00 &  73.877590 & $+$1.024991 & IRAS\,20116$+$3605 & G073.8775$+$01.0245 & Y/N & N & Y & 9.28 & 26 & 25 & -- \\ \hline 
Cl\,137 & 20:16:27.4 & $+$36:54:54 &  74.752565 & $+$0.912439 & IRAS\,20145$+$3645 & G074.7541$+$00.9132 & Y/N & N & Y & 9.29 & 14 & 15 & -- \\ \hline 
Cl\,138 & 21:39:41.1 & $+$51:20:36 &  95.298227 & $-$0.937191 & IRAS\,21379$+$5106 &         --          &  N  & Y & Y &  --  & 21 & 15 &  9 \\ \hline 
Cl\,139 & 21:44:03.1 & $+$55:12:11 &  98.320050 & $+$1.551357 & IRAS\,21423$+$5458 &         --          &  N  & N & Y &  --  &  9 & 10 & -- \\ \hline 
Cl\,140 & 21:49:40.5 & $+$55:24:50 &  99.069107 & $+$1.199380 & IRAS\,21479$+$5510 &         --          &  N  & N & N &  --  & 30 & 15 & -- \\ \hline 
Cl\,141 & 22:35:17.6 & $+$57:59:53 & 105.677118 & $-$0.238204 & IRAS\,22333$+$5744 &         --          &  N  & N & Y &  --  & 29 & 15 & -- \\ \hline 
\multicolumn{14}{|c|}{{\bf Unconfirmed Cluster Candidates}} \\ \hline
Cl\,142 & 04:04:13.4 & $+$51:22:59 & 150.660242 & $-$0.800726 & IRAS\,04004$+$5114 &         --          &  N  & N & Y &  --  & -- & -- & -- \\ \hline 
Cl\,143 & 18:38:16.1 & $-$05:29:33 &  27.419567 & $+$0.866342 & IRAS\,18355$-$0532 &         --          &  N  & N & Y &  --  & -- & -- & -- \\ \hline 
Cl\,144 & 19:22:07.8 & $+$14:29:20 &  49.288593 & $-$0.055411 & IRAS\,19198$+$1423 & G049.2982$-$00.0582 &  N  & N & Y & 5.4  & -- & -- & -- \\ \hline 
\multicolumn{14}{|c|}{{\bf Unclassified Candidate Object}} \\ \hline
Cl\,145 & 16:24:49.9 & $-$46:58:52 & 337.536565 & $+$3.038578 &        --          &         --          &  N  & N & N &  --  & -- & -- & -- \\ \hline 

\end{longtable}
\end{center}
\end{landscape}

\clearpage
\newpage

\section{Cluster Images}\label{app_images}

\subsection{New Clusters}

\begin{center}
\begin{longtable}{|c|p{1.9in}|p{1.9in}|p{1.9in}|}

\caption{\label{new_clusters_images} Images of the newly detected clusters. We
show $JHK(s)$ colour composites from either UGPS or VVV (depending on the
cluster position). The image size is 1\arcmin\,$\times$\,1\arcmin, North is up,
East is to the left. Images are displayed with two different contrasts to show
the various features. The comments column gives a brief description of the
cluster and lists associated objects. The paper with the full Appendix is
available at {\tt\bf http://astro.kent.ac.uk/$\sim$df/papers.html}} \\

\hline \multicolumn{1}{|c|}{\textbf{MHO}} & \multicolumn{2}{c|}{\textbf{Images}} & \multicolumn{1}{c|}{\textbf{Comments}} \\ \hline  \endfirsthead

\multicolumn{4}{c}%
{{\bfseries \tablename\ \thetable{} -- continued from previous page}} \\

\hline \multicolumn{1}{|c|}{\textbf{MHO}} & \multicolumn{2}{c|}{\textbf{Images}} & \multicolumn{1}{c|}{\textbf{Comments}} \\ \hline 
\endhead

\multicolumn{4}{|r|}{{Continued on next page}} \\ \hline
\endfoot

\endlastfoot

\end{longtable}
\end{center}

\subsection{New Cluster Candidates}

\begin{center}
\begin{longtable}{|c|p{1.9in}|p{1.9in}|p{1.9in}|}

\caption{\label{new_cluster_candidates_images} As
Fig.\,\ref{new_clusters_images}, but for the new cluster candidates.} \\

\hline \multicolumn{1}{|c|}{\textbf{MHO}} & \multicolumn{2}{c|}{\textbf{Images}} & \multicolumn{1}{c|}{\textbf{Comments}} \\ \hline  \endfirsthead

\multicolumn{4}{c}%
{{\bfseries \tablename\ \thetable{} -- continued from previous page}} \\

\hline \multicolumn{1}{|c|}{\textbf{MHO}} & \multicolumn{2}{c|}{\textbf{Images}} & \multicolumn{1}{c|}{\textbf{Comments}} \\ \hline 
\endhead

\multicolumn{4}{|r|}{{Continued on next page}} \\ \hline
\endfoot

\endlastfoot

\end{longtable}
\end{center}

\subsection{Known Cluster Candidates}

\begin{center}
\begin{longtable}{|c|p{1.9in}|p{1.9in}|p{1.9in}|}

\caption{\label{known_cluster_candidates_images} As
Fig.\,\ref{new_clusters_images}, but for the known cluster candidates.} \\

\hline \multicolumn{1}{|c|}{\textbf{MHO}} & \multicolumn{2}{c|}{\textbf{Images}} & \multicolumn{1}{c|}{\textbf{Comments}} \\ \hline  \endfirsthead

\multicolumn{4}{c}%
{{\bfseries \tablename\ \thetable{} -- continued from previous page}} \\

\hline \multicolumn{1}{|c|}{\textbf{MHO}} & \multicolumn{2}{c|}{\textbf{Images}} & \multicolumn{1}{c|}{\textbf{Comments}} \\ \hline 
\endhead

\multicolumn{4}{|r|}{{Continued on next page}} \\ \hline
\endfoot

\endlastfoot

\end{longtable}
\end{center}

\subsection{Unknown Object}

\begin{center}
\begin{longtable}{|c|p{1.9in}|p{1.9in}|p{1.9in}|}

\caption{\label{unknown_object_images} As Fig.\,\ref{new_clusters_images}, but
for the object that cannot be classified.} \\

\hline \multicolumn{1}{|c|}{\textbf{MHO}} & \multicolumn{2}{c|}{\textbf{Images}} & \multicolumn{1}{c|}{\textbf{Comments}} \\ \hline  \endfirsthead

\multicolumn{4}{c}%
{{\bfseries \tablename\ \thetable{} -- continued from previous page}} \\

\hline \multicolumn{1}{|c|}{\textbf{MHO}} & \multicolumn{2}{c|}{\textbf{Images}} & \multicolumn{1}{c|}{\textbf{Comments}} \\ \hline 
\endhead

\multicolumn{4}{|r|}{{Continued on next page}} \\ \hline
\endfoot

\endlastfoot

 %
 %

\end{longtable}

\label{lastpage}
\end{center}

\end{appendix}

\end{document}